\documentclass[]{spie}  
\usepackage[]{graphicx}
\usepackage{float}
\usepackage{wrapfig}
\usepackage{subfig}

\title{ProtoEXIST: Advanced Prototype CZT Coded Aperture Telescopes for EXIST} 

\author{Branden Allen$^*$\supit{a}, Jaesub Hong\supit{a}, Josh Grindlay\supit{a},\\
	Scott D. Barthelmy\supit{b}, Robert G. Baker\supit{b}, Neil A. Gehrels\supit{b},\\
	Trey Garson\supit{c}, Henric S. Krawwczynski\supit{c},\\
	Walter R. Cook\supit{d}, Fiona A.Harrison\supit{d},\\
	Jeffery A. Apple\supit{e}, Brian D. Ramsey\supit{e}
\skiplinehalf
\supit{a}Harvard-Smithsonian CfA, 60 Garden Street, Cambridge, MA 02138, USA\\
\supit{b}NASA Goddard Space Flight Center, Greenbelt, MD 20771, USA\\
\supit{c}Washington University, St. Louis, MO, 63130, USA\\
\supit{d}California Institute of Technology, Pasadena, CA 91125, USA\\
\supit{e}NASA Marshall Space Flight Center, VP62, Huntsville, AL 35812, USA\\
}

\authorinfo{$^*$Send correspondence to Branden Allen (ballen@cfa.harvard.edu)}

\begin{document} 
\maketitle 

\begin{abstract}
{\it ProtoEXIST1} is a pathfinder for the {\it EXIST-HET}, a coded 
aperture hard X-ray telescope with a 4.5 m$^2$ CZT detector plane 
a 90$\times$70 degree field of view to be flown as the primary instrument on the 
{\it EXIST} mission and is intended to monitor the full sky every 3 h in an 
effort to locate GRBs and other high energy transients.  
{\it ProtoEXIST1} consists of a 256 cm$^2$ tiled CZT detector plane containing 4096 
pixels composed of an 8$\times$8 array of individual 1.95 cm $\times$ 1.95 cm 
$\times$ 0.5 cm CZT detector modules each with a 8 $\times$ 8 pixilated anode 
configured as a coded aperture telescope with a fully coded $10^\circ\times10^\circ$ 
field of view employing passive side shielding and an active 
CsI anti-coincidence rear shield, recently completed its maiden flight out
of Ft. Sumner, NM on the 9th of October 2009.  During the duration of its 
6 hour flight on-board calibration of the detector plane was carried out utilizing a 
single tagged 198.8 nCi Am-241 source along with the simultaneous 
measurement of the background spectrum and an observation of Cygnus X-1.
Here we recount the events of the flight and report on the detector performance 
in a near space environment.  We also briefly discuss {\it ProtoEXIST2}: the next stage of 
detector development which employs the {\it NuSTAR} ASIC enabling finer 
(32$\times$32) anode pixilation. When completed {\it ProtoEXIST2} will consist of a 256 cm$^2$ 
tiled array and be flown simultaneously with the ProtoEXIST1 telescope.
\end{abstract}


\section{Introduction}\label{sec:intro}
CZT (CdZnTe Cadmium-Zinc-Telluride) and CdTe (Cadmium-Telluide) is a semiconductor
which has found increasing use in hard X-ray and soft $\gamma$-ray sensing applications 
over the past decade.  The wide band gap ($\sim2.5$ eV) and high Z-composition make the 
detection of high energy X-rays / soft $\gamma$-rays ($\sim$5-2000 keV) possible at room 
temperature and eliminates the need for cryogenics which is traditionally required 
for other commonly utilized semiconductor detector materials such as high purity germanium 
(bandgap $\sim0.62$ eV) or silicon ($\sim1.1$ eV).
For application in X-ray astronomy this removes a significant constraint to
operation in a space environment through appreciable mass savings
and eliminates the need for irreplaceable consumables increasing the upper bound of 
the mission lifetime.  Long term implementation of high energy coded aperture
telescopes utilizing CZT and CdTe detector planes have in fact been successfully 
demonstrated over the past decade in the {\it INTEGRAL-IBIS}\footnote[2]{Launch Date:
17$^\mathrm{th}$ October 2002 (Baikonur)} (Imager on Board {\it INTEGRAL}) using 
CdTe\cite{2003A&A...411L.131U,2003A&A...411L.141L} 
and the {\it SWIFT-BAT}\footnote[3]{Launch Date: 20$^\mathrm{th}$ November 2004 (Cape Canaveral)} 
(Burst alert telescope) with CZT\cite{2005SSRv..120..143B}.  The success of 
{\it INTEGRAL} and {\it SWIFT} has been encouraging and CZT/CdTe is finding 
increased use in astronomy.  In this first generation of space based detectors the 
individual pixels consisted of an individual detector crystal with a single readout.  
Advances in CZT detector technology have long since shown that the partitioning 
of a single detector into multiple detector pixels is possible by application 
of a pixilated anode without degradation in detector performance 
\cite{1995PhRvL..75..156B, 1994NIMPA.353..356D}.  This advance coupled with the 
introduction of new readout ASICs (application specific integrated chips) which 
consume less power-per-channel will enable the creation of even larger CZT 
detector planes at lower cost.  Indeed such pixilated detectors have been
flown in limited quantities as part of balloon payloads e.g. HEFT\cite{2005ExA....20..131H} 
with a single detector and InFOC$\mu$S\cite{2003SPIE.4851..945B} with 4 detectors in 
a $2\times2$ configuration.

The {\it ProtoEXIST} program was conceived as a pathfinder for the development of 
an instrument with such a detector plane on a much larger scale: the 
{\it EXIST-HET}\cite{2009SPIE.7435E...6H,2010THIS...00...000H} (high energy telescope on {\it EXIST}).  
The {\it HET} is currently envisioned to utilize 4.5 m$^2$ of CZT consisting of 
11264 20 mm $\times$ 20 mm $\times$ 5 mm pixilated CZT detector 
units with 12$\times10^6$ pixels total.  Each detector units will have 32 $\times$ 32
pixels with a 600 $\mu$m pixel pitch and consume 20 $\mu$W per channel.  In order to 
demonstrate the feasibility of such a detector plane using a scalable and highly redundant 
architecture the {\it ProtoEXIST} project has been divided into three phases:
\begin{enumerate}
	\item {\it ProtoEXIST1} (described here) is a coded-aperture telescope with a 
	256 cm$^{2}$ CZT detector plane tested in a space environment with a balloon 
	flight at an altitude of $\sim40$ km above mean sea level.  The detector plane 
	utilizes the {\it RADNET} ASIC and is comprised of an $8\times8$ array of individual 
	CZT detectors with pixilated anodes.  Each individual detector had 64 pixels 
	(a $8\times8$ array of pixels) or 4096 individual pixels over the entire detector plane.
	This phase culminated in a 6 hour flight out of Ft. Sumner, NM on the 9$\mathrm{th}$
	of October 2009.
	\item {\it ProtoEXIST2} will demonstrate the feasibility of a nearly identical 
	architecture utilizing the {\it NuSTAR} DB-ASIC (i.e. direct bond ASIC)
	\cite{2005ExA....20..131H,2009SPIE.7435E...2R}.  This enables the demonstration 
	of a detector plane with pixels possessing a 600 $\mu$m pitch 
	(i.e. a $32\times32$ pixel array for each detector) and fulfills the fine 
	pixilization requirement of the {\it EXIST-HET}.  The DB-ASIC currently consumes 
	80 $\mu$W per channel exceeding the {\it EXIST} power requirement, 
	necessitating the production of a new lower power version.
	Currently the design of the required detector electronics is nearing completion.
	At the time of this writing it is expected that testing of the first detectors 
	of this series will begin within the next few months (i.e. October 2010).
	\item {\it ProtoEXIST3} will demonstrate the final unmet requirement for {\it EXIST}
	through a relatively minor modification of the DB-ASIC in order to produce the 
	EX-ASIC ({\it EXIST}-ASIC) which will consume 20 $\mu$W per 
	channel.  This phase of the program will conclude with a flight of the original 
	detector module from {\it ProtoEXIST1} and a second detector module consisting 
	of 1/4 DB-ASIC mounted CZT detectors and 3/4 EX-ASIC mounted CZT detectors.
\end{enumerate}

\section{ProtoEXIST1 Payload Overview}\label{sec:payload}
The {\it ProtoEXIST1} payload consists of two main parts: the gondola which supports
flight operations and pointing functions, and the instrument package which
contains all required subsystems for the support of the {\it ProtoEXIST1} coded aperture
telescope.

\begin{figure}
	\centering
	\includegraphics[width=\textwidth]{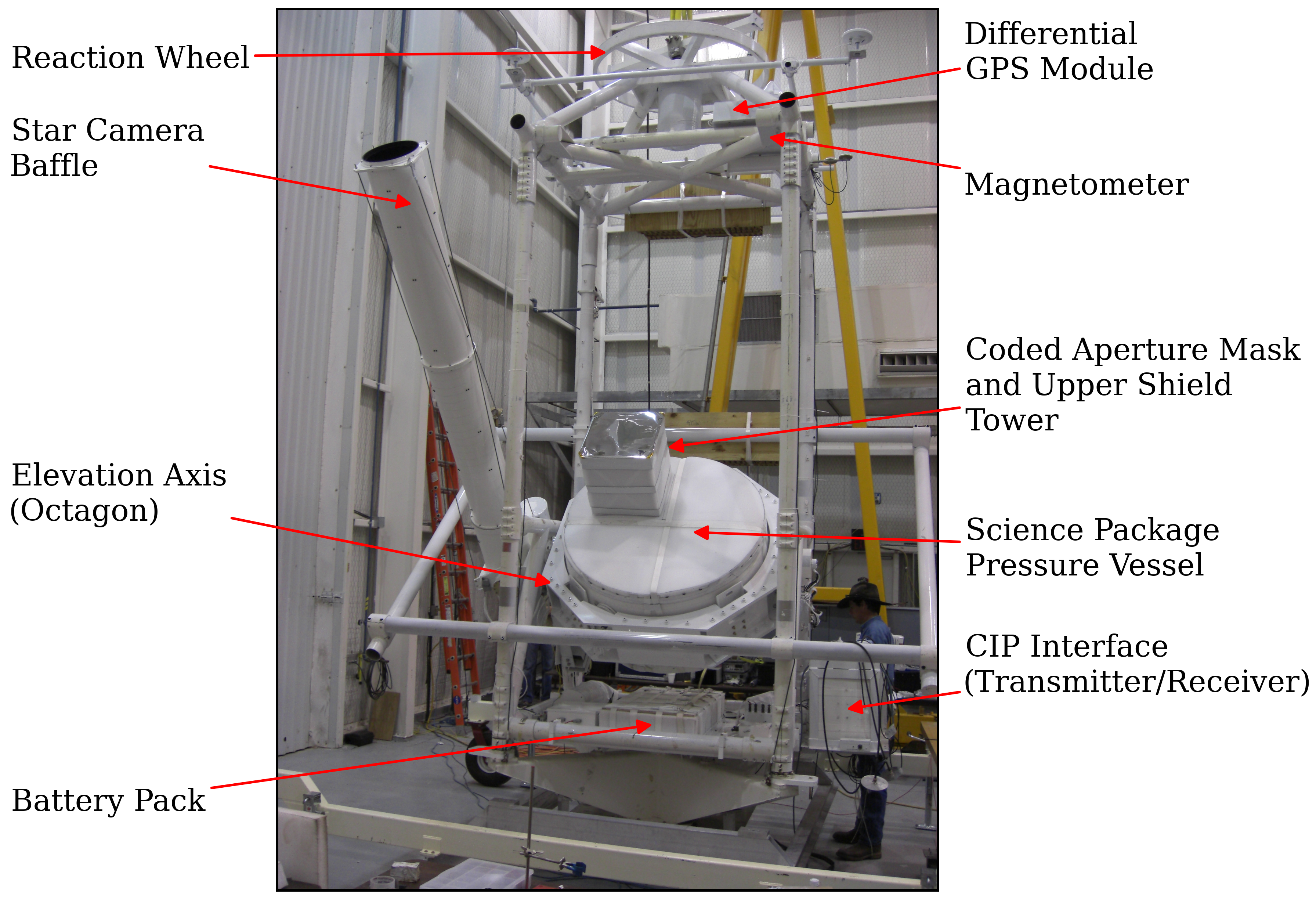}
	\caption{The integrated {\it ProtoEXIST1} gondola 5 days prior
	to flight.  The gondola supports the pointing, tracking, communications and
	timing subsystems and has an approximate height of 5 m.  The science package
	pressure vessel is directly mounted to the gondola through its octagonal support
	frame, which serves as the telescope's elevation axis.  Visible to the left of the 
	pressure vessel is the star camera used for independent aspect determination, which
	is mounted directly to the same elevation axis to preserve the boresight between 
	the X-ray telescope and the star camera.}
	\label{fig:gondola}
\end{figure}

The main {\it ProtoEXIST1} gondola (pictured in figure \ref{fig:gondola}) was refurbished using 
the existing {\it EXITE} \cite{1986ITNS...33..750G} gondola frame which flew multiple 
times during the late 1980s and early 1990s.  The pointing system as well as
all supporting electronic subsystems were replaced using the {\it HERO}\cite{2002ApJ...568..432R} 
(High Energy Replicated Optics) gondola support electronics and pointing system design
as a model. As in {\it HERO} the coarse azimuthal pointing system is based on a 
differential global positioning system (DGPS) unit with a backup magnetometer in 
place in the event of DGPS failure. A daytime star camera system (also based on 
a design from {\it HERO}\cite{2002OptEn..41.2641D}) and inertial 
guidance system are utilized for fine pointing control.  The star camera has a 
$3^\circ\times2^\circ$ field of view with a small pixel size ($3.5^{''}\times3.5^{''}$) which 
together with a long ($\sim2.5$ m) baffle enables the resolution of stars over the 
daytime sky background.  In addition two transmitters are mounted to the gondola, one for the 
transmission of reduced data to the ground for storage and analysis, and the other 
for the monitoring of the gondola and payload status.  A single receiver is present 
for command and control and is linked to the standard command interface package (CIP) 
provided for scientific balloon flights by the Columbia scientific balloon facility (CSBF)
\footnote[2]{http://www.csbf.nasa.gov}.  The gondola also provides universal time (UT) 
synchronization information to the entire payload and experiment using a separate GPS receiver.
\begin{figure}
	\centering
	\includegraphics[width=\textwidth]{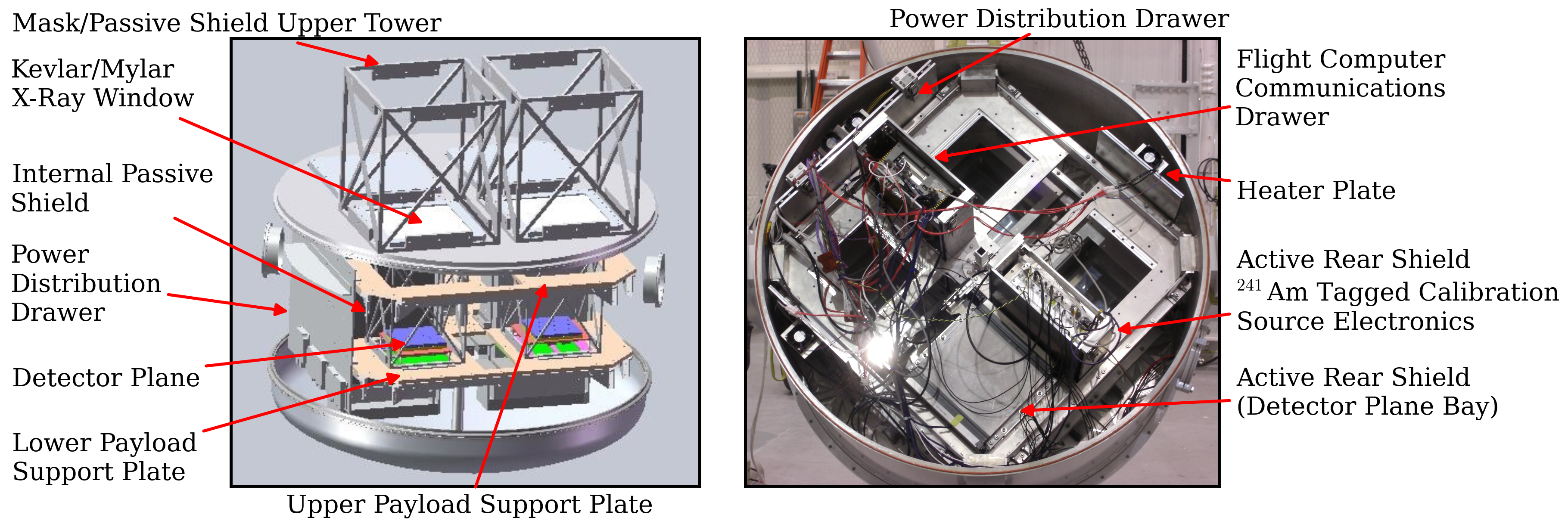}
	\caption{{\bf Left}: A 3D model of the internal structure of the main science 
	package pressure vessel.  The vessel is shown here with two telescopes
	integrated into the payload, however only one full detector was produced and
	integrated for the flight. {\bf Right}: a view of the rear of the fully integrate
	science package.  The single integrated telescope sits directly above the active 
	rear shield visible in the lower telescope port.}
	\label{fig:vessel}
\end{figure}

The remainder of the payload is contained within or mounted to a single 1.252 m
diameter cylindrical aluminum pressure vessel with two removable end caps for payload 
access, and two removal self-sealing connector flanges which provide the electrical 
interface between the vessel and the gondola.  For {\it ProtoEXIST1} a single telescope
consisting of: 
\begin{enumerate}
	\item the CZT detector plane (\S\ref{sec:det})
	\item a lower (internal to the pressure vessel) side shield (graded Z-shield) (\S\ref{ssec:shld})
	\item a upper side shield and mask support structure (\S\ref{ssec:shld})
	\item the active CsI rear shield (\S\ref{ssec:rshld})
	\item a single 198.9 nCi tagged $^{241}$Am calibration source (\S\ref{ssec:rshld})
	\item the coded aperture mask (\S\ref{ssec:mask})
\end{enumerate}
is mounted in one of four available telescope slots, each of which is capable of
accommodating a detector plane mounted within a EMI/RFI shielded enclosure with an approximate 
maximum size of 20 cm $\times$ 20 cm.  The shielded enclosure has a thin (6 mil) Al X-ray window
for the observation of external sources. The internal structure of the pressure vessel 
(represented by the 3D model in figure \ref{fig:vessel}) contains two main component mounting plates 
which are oriented perpendicular to the axis of the main vessel 
cylinder and are separated from one another by a distance of 288 mm. 
The mounting plate located nearest to the bottom cap serves as the optical bench 
for the detector planes and as the mount point for the active rear shield 
(see \S\ref{ssec:rshld}).  The top surface also contains mount points for the internal 
half of the passive side shielding (see \S\ref{ssec:shld}) which surrounds the detector 
plane and runs up to the bottom surface of the top cap.
The flat pressure vessel top cap is located approximately 400 cm above the top surface 
of the detector plane and is nominally 25.4 mm thick.  The inner surface of the top cap 
has a honeycomb structure with a pocket depth of 14.48 mm to minimize distortion of the top 
cap due to internal pressure during flight while 
reducing the overall weight by about 30\%.  4 ports are cut into the top cap onto 
which an airtight low mass X-ray window is mounted through which the detector 
plane is exposed to the overhead sky.  The transparent region of the X-ray window 
assembly measures 28 cm $\times$ 28 cm and is composed of a external 0.61 mm 
thick Kevlar layer and a single internal 0.12 mm thick mylar layer attached to an 
aluminum mounting frame.  The Kevlar provides mechanical strength against the internal 
vessel pressure while the mylar provides the airtight seal.  The thin, light design 
enables transmission of X-rays down to approximately 5 keV.  Mounting points are located 
on the top surface of the pressure vessel top cap for the superstructure which supports
coded aperture mask which in the final integrated payload is located 513 mm above the 
upper surface of the top cap.

Additional space within the vessel adjacent to the detector plane mounting ports serve 
as mount points for the science package electrical subsystems.  
There are eight additional ports in all: 4 centrally located between detector 
planes and 4 located in the area between the mounting plate and the outer vessel skin.
In the current ({\it ProtoEXIST1}) configuration a flight computer, a transmitter 
interface card, 3 Ethernet controlled heater control relays, the active shield 
and tagged $^{241}$Am calibration source control box, 2 primary pressure vessel heating 
elements, a power control, distribution, and monitoring box, and an Ethernet switch which 
serves as the hub of the science package internal communication system 
are mounted along with the prototype detector plane inside the vessel 
(see figure \ref{fig:vessel}).  The flight computer is the primary command/control and 
monitoring device for the science package and oversees all internal operations 
with the exception of the primary power control.  The individual payload components 
communicate via an internal Ethernet network which serves as the primary communications 
link between the various subsystems.  The flight computer is connected with a serial port 
(RS-422) to the receiver on the gondola which serves as the command uplink for the 
science package.  A command interpreter running continuously on the flight computer 
monitors the uplink for incoming commands and, when a command is received, determines
the required action to be taken.  The command is then (in most cases) forwarded across the 
internal network to the relevant subsystem.  The flight computer is also linked to a transmitter
interface card using a RS-422 interface and also acts as the primary downlink.  The downlink
is used to monitor the status of the instrument package and its associated subsystems
as well as transmit reduced data packets from the detector plane to the ground.
Each subsystem within the pressure vessel, including the detector plane, transmits status 
information over the network back to the flight computer where the downlink monitor, another
persistent process running on the flight computer, monitors the incoming data, collates it 
and transmits all information back to the ground station.  In order to prevent deadlocks
the power monitor and distribution subsystem receives commands directly from the CIP 
interface, however status information is still sent via the flight computer to the downlink.

The entire science package is physically supported by the mounting plates which are coupled 
to 8 hard points embedded in the pressure vessel main body skin which in turn are attached 
directly to the octagonal support structure (visible surrounding the vessel in figure 
\ref{fig:gondola}).  This serves simultaneously as the main load bearing structure for 
the internal mechanical structure of the vessel and as the elevation axis for 
the coded-aperture telescope.  The octagonal support structure mounts directly 
to the {\it ProtoEXIST1} gondola via two elevation flanges which rotate on bearings 
affixed to the gondola superstructure.  The elevation is monitored using a shaft
angle encoder (SAE) located on the flange.

\section{The {\it\bf ProtoEXIST1} Detector Plane}\label{sec:det}
The {\it ProtoEXIST1} detector plane utilizes a modular architecture for
redundancy and scalability. This design (as outlined section \ref{sec:intro}) 
is to be utilized for the construction of a much larger array i.e.{\it EXIST}.  
The {\it ProtoEXIST1} detector plane consists of three primary components: 
a single flight control board (FCB), 8 detector crystal arrays (DCAs), 
64 detector crystal units (DCUs)\cite{2009NIMPA.605..364H,2007SPIE.6706E...8H,2006SPIE.6319E..23H}.
\begin{figure}
	\centering
	\includegraphics[width=\textwidth]{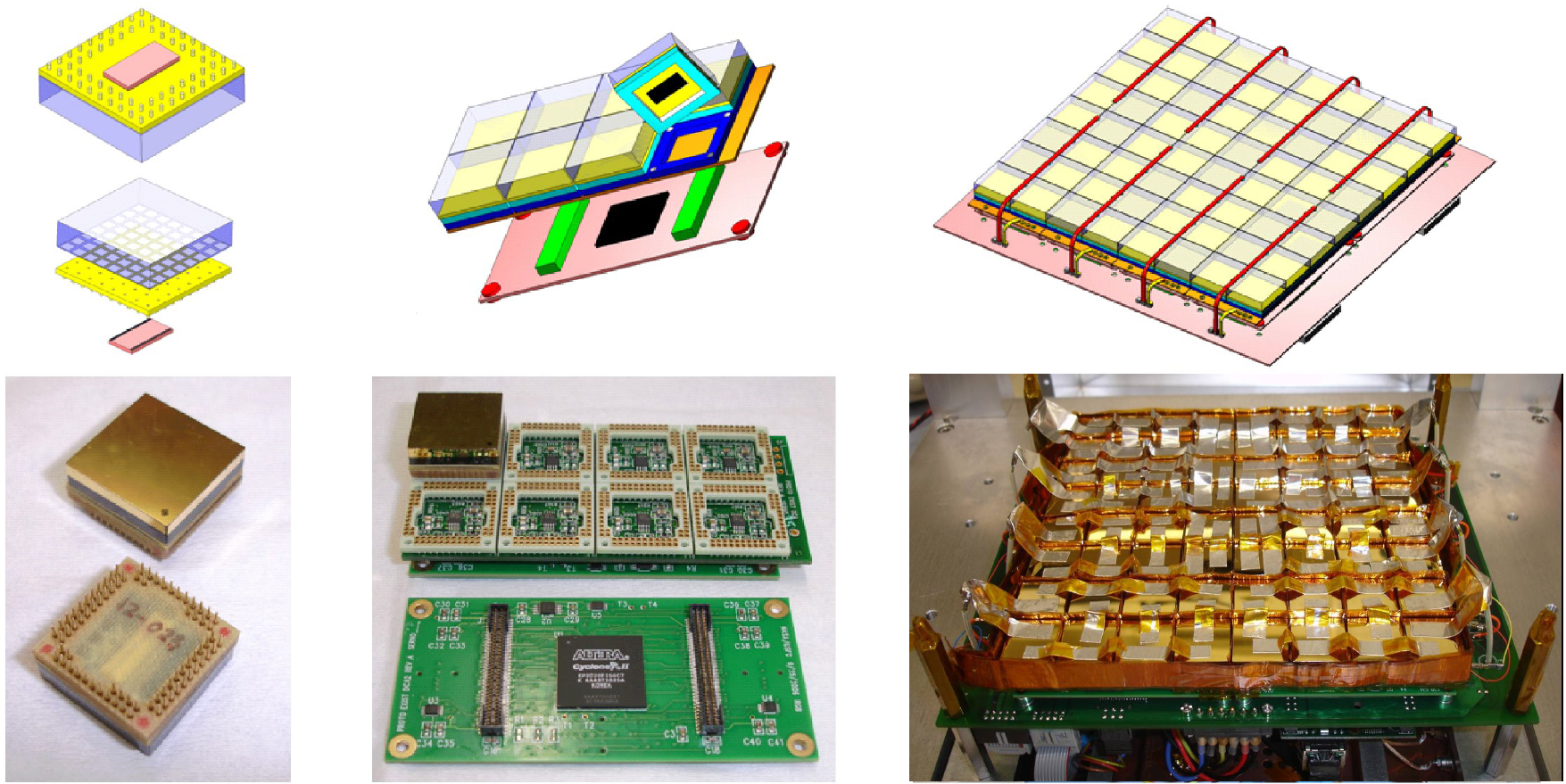}
	\caption{{\bf Left:} A single 19.5 mm $\times$ 19.5 mm $\times$ 5 mm CZT crystal 
	(shown in purple) bonded to the top surface of an interposer board (IPB, shown in yellow) 
	constitutes a 
	single detector crystal unit (DCU).  The RadNet ASIC (shown in red) which reads out 
	events and measures deposited charge on the CZT is bonded to the underside of the IPB.
	{\bf Center:} A single detector crystal array (DCA) pictured with 
	a single DCU mounted, supports a $4\times2$ array of DCUs ($\sim32$ cm$^2$) .  
	The lower DCA board contains a single FPGA which serves as the sole link for 
	data transmission and command input for the mounted DCUs.
	{\bf Right:} The fully integrated {\it ProtoEXISTI} detector plane with the detector 
	module cover removed.  A $2\times4$ array of DCAs ($\sim256$ cm$^2$) is mounted to 
	the top surface of a flight control board (FCB) and constitutes a single 
	complete detector plane.   The individual detectors are coupled to a high 
	voltage bias (nominally -600 V) through conductive Al tape (visible above) 
	which are coupled to eight individual leads that are in turn connected to 
	two high voltage power supplies located on the FCB.}
	\label{fig:det}
\end{figure}

As shown in figure \ref{fig:det} each DCU consists of a 
19.5 mm $\times$ 19.5 mm $\times$ 5 mm CZT crystal with a 8 $\times$ 8 
pixilated anode bonded to a single interposer board (IPB) with single RadNet ASIC
bonded to the side opposite the CZT.  The CZT crystals are procured from Redlen technologies
\footnote[1]{www.redlen.com} which pioneered the mass production of high quality low
cost CZT using the traveling heater method (THM). CZT thusfar received have typically exhibited
leakage currents below 0.6 nA per pixel and are well within the operational specs of the
ASIC (c.f. Allen {\it et. al.} in preparation\cite{2010FUTU...00...000A}). The electrical and mechanical 
connection of the CZT is achieved using a low temperature solder bond\footnote[2]{Low temperature solder
bonding carried out at Delphon/QuikPak (http://www.delphon.com)} 
($\sim40\%$ of all {\it ProtoEXIST1} detectors 
constructed) or transient liquid phase sintering (TLPS) bond\footnote[3]{TLPS bonding 
carried out at Creative Electron (http://www.creativeelectron.com)} between each CZT 
anode pixel and the corresponding 
IPB pad.  The IPBs contains traces which route connections from the anode pads on 
the upper surface to output pads located on the underside of the IPB.  The final
connection of each pixel to the ASIC is made with a wirebond
between the output pad and the ASIC input pad.  A single  square multi-pin connector 
is located on the underside of the ASIC which mates to a socket located on the upper 
side of the detector crystal array (DCA) upper board.

The DCA is composed of two printed circuit boards (PCB) with matching dimensions. Located
on the upper surface of the upper board are $4\times2$ sockets to which individual DCUs 
are mounted to form a approximately 4 cm $\times$ 8 cm (32 cm$^2$) detector module.  The upper board 
also contains analog-to-digital converters (ADC) for the digitization of the ASIC sample 
readout as well as digital-to-analog converters (DAC) for 
the input of settings into the ASIC (i.e. thresholds, etc.).  The bottom surface of the 
board supports two connectors which make the connection to the lower DCA board.  The lower
DCA board contains a single FPGA which handles the processing of data as well as command
and control of each of the 8 DCUs mounted to the DCA.  In order to reduce the number of 
required lines in the board the DCA FPGA handles 4 independent data channels each of 
which consists of the readout from 2 DCUs (a single DCU pair).  Each DCU pair is effectively
treated as a single detector except that threshold for each DCU may still be set independently.
The bottom surface of the lower DCA board contains three surface mount connectors which 
interface to the flight control board (FCB).

The FCB supports a $2\times4$ array of DCAs (a total of 64 individual DCUs) which constitutes
the complete 16 cm $\times$ 16 cm (256 cm$^2$) detector plane.  Two EMCO HVPSs are mounted to 
the underside the FCB, each with 4 leads (one per DCA) mounted on opposite sides of the detector plane.  
The HV leads and the cathodes of the individual DCUs are coupled with conductive Al tape 
(see figure \ref{fig:det}).  A thin flexible RFI/EMI shield consisting of a copper ground 
plane sandwiched between two layers of kapton tape surrounds the individual DCU to 
eliminate cross-talk between neighboring DCUs as well as reduce in the influence of any external
sources of noise (c.f. Allen {\it et. al.} in preparation\cite{2010FUTU...00...000A}).
For handling of the data as well as command and control functions the FCB contains a single 
FPGA.  All 32 (4 from each DCA) independent data channels are routed into the FCB-FPGA and 
combined into a single data stream.  The FPGA also serves as 
the terminal input point for timing, anti-coincidence, and calibration source signals.  
The FCB has 8 TTL inputs to accommodate discriminator output from the CsI rear shield, 
the $^{241}$Am tagged calibration source, and the 1 PPS time synchronization signal from 
the gondola GPS clock.  In the {\it ProtoEXIST1} flight configuration four of these inputs 
were utilized: two were required for each PMT on the CsI rear shield, and a single input was 
required for both the calibration source the 1 PPS GPS input.  All of these signals are routed
into the FCB FPGA where tagging and timing information may be applied to the event data as
it is received.  To ensure that the correct events are assigned the correct timing/tagging 
flags each of the 32 data channels from the detector pairs supports a fast trigger line, 
which connects directly to the FPGA.  The fast trigger signals in combination with the 
external discriminator inputs are compared with an internally set 5 $\mu$s 
anti-coincidence window to identify background and calibration source events in the 
detector.  After application of timing and tagging information the single data stream 
is sent from the FCB-FPGA to a Coldfire MCF5282 microcontroller\footnote[4]{http://www.freescale.com}
mounted on a Mod 5282 NetBurner card
\footnote[5]{http://www.netburner.com/products/core\_modules/mod5282.html}.  
The NetBurner card implements an Ethernet interface with the MCF5282 microcontroller 
and serves as the FCB's primary interface with the outside world.  The MCF5282 serves
4 functions: the organization of the data stream into packets for transmission over
Ethernet, the monitoring of the Ethernet port for user commands and the implementation
of those commands on the proper subsystem, control of the high voltage supplies for the 
biasing detectors on the FCB, and the monitoring of the detector plane housekeeping data.
The housekeeping data contains general status information on voltages, currents, and 
the board temperature which is combined into independent data packets and transmitted over the Ethernet.
In {\it ProtoEXIST1} all data and housekeeping packets are transmitted to the flight
computer for storage on the payload data recorder (a solid state hard drive contained within 
the flight computer) and simultaneous downlink (see \S\ref{sec:payload})

\section{Coded Aperture Telescope}\label{sec:tel}
The {\it ProtoEXIST1} telescope occupies a single bay (out of a total of 4)
in the pressure vessel and consists of four primary components: the graded Z-shield 
(\S\ref{ssec:shld}) which restricts the detector plane's field of view to the coded 
aperture mask and partially shields the it from atmospheric X-ray albedo, the CsI 
active rear shield (\S\ref{ssec:rshld}) which enables the tagging atmospheric X-ray 
albedo events as well as the identification of cosmic ray background.
The coded aperture mask (\S\ref{ssec:mask}), a perfect binary uniform redundant array
(URA) which is mounted 900 mm away from the cathode surface of the 
detector plane and enables the reconstruction of hard X-ray source positions, and
the pixilated 16 cm $\times$ 16 cm position sensitive detector plane (\S\ref{sec:det}) 
described above.

\subsection{Graded Z-Shield}\label{ssec:shld}
As previously alluded to the graded Z-shield is a passive shield installed in 
order to restrict the detector plane field of view to the coded aperture mask
and to minimize the effects of the atmospheric albedo.  The shield for a single telescope
is divided into a 26 cm $\times$ 26 cm $\times$ 39.7 cm lower section, 
which sits inside the pressure vessel and surrounds the detector plane on 4 sides, 
and a 40 cm $\times$ 40 cm $\times$ 50 cm upper section which is affixed to the 
top cap of the science package pressure vessel.  Both sections of the side shield 
are composed of laminated sheets of Pb-Sn-Cu (ordered from the outermost to the 
innermost layer) attached to an aluminum support frame.  The design of the side 
shielding is based on the {\it SWIFT-BAT} graded Z-shield\cite{2003SPIE.4851.1374R}, 
however {\it SWIFT-BAT} employed a Z-shield composed of Pb-Sn-Ta-Cu.  Due to the 
prohibitive cost of Ta at the time of construction this layer was dropped from the 
{\it ProtoEXIST1} design and the thickness of the Sn and Cu layers were increased 
in order to compensate for its removal.
\begin{table}
	\centering
	\include{zshield}
	\caption{A list of the minimum required thickness for the individual layers of the 
	{\it ProtoEXIST1} graded Z-shield.  The actual thickness of the final structures measured
	are given in parentheses.}
	\label{tbl:shld}
\end{table}
The inner layers of the graded Z-shield serve to attenuate the Pb fluorescence lines and, 
similarly, the innermost Cu layer attenuates the remaining Sn fluorescence lines,
the Cu fluorescence lines ($\sim8$ keV) lie below the minimum detector threshold.  
Due to the discovery of a minor mechanical problem with the internal vessel structure a 
rebuild of the lower section of the side shield was required.  As a result a portion 
of the Pb layer at the top of the lower section approximately 30 cm distant from the detector 
plane surface was left unshielded due the lack of additional Sn and Cu in the field.  
As a result the Pb-K$\alpha$ and -K$\beta$ florescence lines were observed emanating from 
the exposed section during the flight (see \S\ref{sec:flt}).

\subsection{Active Rear Shielding and $^{241}$Am Calibration Source}\label{ssec:rshld}
The shield electronics box is the subsystem/module that is responsible for the readout
of raw PMT signals from the active rear shield and the $^{241}$Am calibration source.
For this purpose the shield electronics box contains HV outputs for the biasing of PMTs,
and computer controlled discriminator thresholds and outputs.  The shield system is 
controlled by the flight computer over a serial port.  A connection to the internal 
Ethernet network is available and used to send raw PMT pulse rates for both
the CsI rear shield and the $^{241}$Am source.  The discriminator outputs for each
PMT are linked directly to the detector plane for anti-coincidence timing.
The active rear shield itself is composed of a single 26 cm $\times$ 26 cm $\times$ 2 cm 
CsI crystal coupled to two PMTs across two wave shifter bars mounted on opposite sides
of the scintillator.  The PMTs are monitored individually and have separate discriminator
output lines. The calibration source consists of a small aluminium cylinder that is filled with a 
$^{241}$Am doped scintillator.  A clear window to which a single PMT is optically coupled 
is located at the top end of the cylinder which enables the detection 5.47 MeV $\alpha$-particles 
which are emitted in coincidence with the 59.6 keV X-rays used for detector calibration.
The calibration source PMT has an independent discriminator input/output line on the 
shield electronics box as well that enables the identification or tagging of calibration
X-ray events in the detector plane.  The calibration source tagging efficiency observed in 
the hanger and during the flight is approximately 75\%.

\subsection{Detector Integration and Performance}\label{ssec:det_per}
The integration of the {\it ProtoEXIST1} proceeded in two steps.  Newly bonded DCUs 
were received at a rate of about 8-10 per week during peak production 
and integrated onto a single DCA for preliminary testing.  Due to the production rate 
and time constraints most DCUs were not tested individually but tested together in a 
fully populated DCA.  Previous tests have not revealed a significant difference in 
performance for healthy detectors operating in an array provided that the EMI/RFI 
shielding completely surrounds each
DCU and is properly grounded.  The test array would then undergo a short pulser test
at 0 V for the detection of dead pixels, a pulser test at 600 V for the detection of
hot pixels and any other possible anomalies, and finally undergo a $^{57}$Co calibration
test.  Detectors with more than 3 dead pixels or other anomalies were replaced and
the DCA retested as needed.  Once a good candidate DCA was completed it was set aside
for integration into the flight detector.

The flight detector was then integrated 1 DCA at a time.  After the installation a single
DCA the 0 V pulser, 600 V pulser, and $^{57}$Co calibration tests were repeated on the partially
completed array to detect any changes with the addition of a single DCA.  This was 
repeated until the installation of all 8 fully populated DCAs was complete.  During 
the detector integration and after arrival in the field at Ft. Sumner some detectors 
$\sim5$ developed anomalies and required replacement.  

During each integration step, with the exception of a few anomalous detectors which 
went bad during the course of pre-flight testing, no substantial changes in detector 
performance were observed between the operation individual isolated DCUs, 
DCUs integrated into a single $2\times4$ array on a DCA, and DCAs integrated 
into the array of $2\times4$ DCAs ($64\times64$ DCUs).  In each case the detectors 
maintained an average resolution of 3.5 keV (FWHM) at 59.6 keV.
A more complete account will be available in Allen {\it et. al.}\cite{2010FUTU...00...000A}.

\subsection{Coded Aperture Mask and Imaging Performance}\label{ssec:mask}
The {\it ProtoEXIST1} telescope utilizes a $64\times64$ element prefect binary uniform 
redundant array (URA) (c.f. references\cite{1978ApOpt..17..337F,1998ExA.....8...97B}) 
placed 900 mm from the surface of the detector plane for image reconstruction.
4.2 mm square pixels with a pitch of 4.7 mm are utilized for the mask pattern, 
the support grid (i.e. the material for support of the mask structure running between pixels) 
has a thickness of 0.5 mm.  The mask is composed of 12 identical laminated tungsten sheets
into which the URA pattern was chemically etched
\footnote[1]{Photo-etched at Tech-Etch, Plymouth, MA http://www.tech-etch.com}.  
The assembled mask is mounted into an aluminium frame which sits atop the outer 
side shield tower on the upper surface of the {\it ProtoEXIST1} pressure vessel.  
Due to the difference between the coefficients of thermal expansion for Al and W 
($\alpha_\mathrm{Al}/\alpha_\mathrm{W}\sim5$) and the wide range of temperatures 
experienced in the upper atmosphere (between $-70^\circ$C and $20^\circ$C), the mask is coupled to the 
mounting frame via a Si rubber gasket in order to prevent damage or warping of the mask 
pattern as the frame expands and contracts during the flight.  The mask also has 8 hard
attachment points (at 1 each corner and 1 at each midpoint) buffered by a gasket to 
ensure that the mask remains in the same position relative to the detector plane during 
the flight.  Four additional 3 mm diameter holes are etched into each pattern and are used 
for the attachment of alignment pins which prevent the mask layers from slipping out of
alignment.
\begin{figure}
	\centering
	\subfloat[]{\label{sfig:mask}\includegraphics[width=0.29\textwidth]{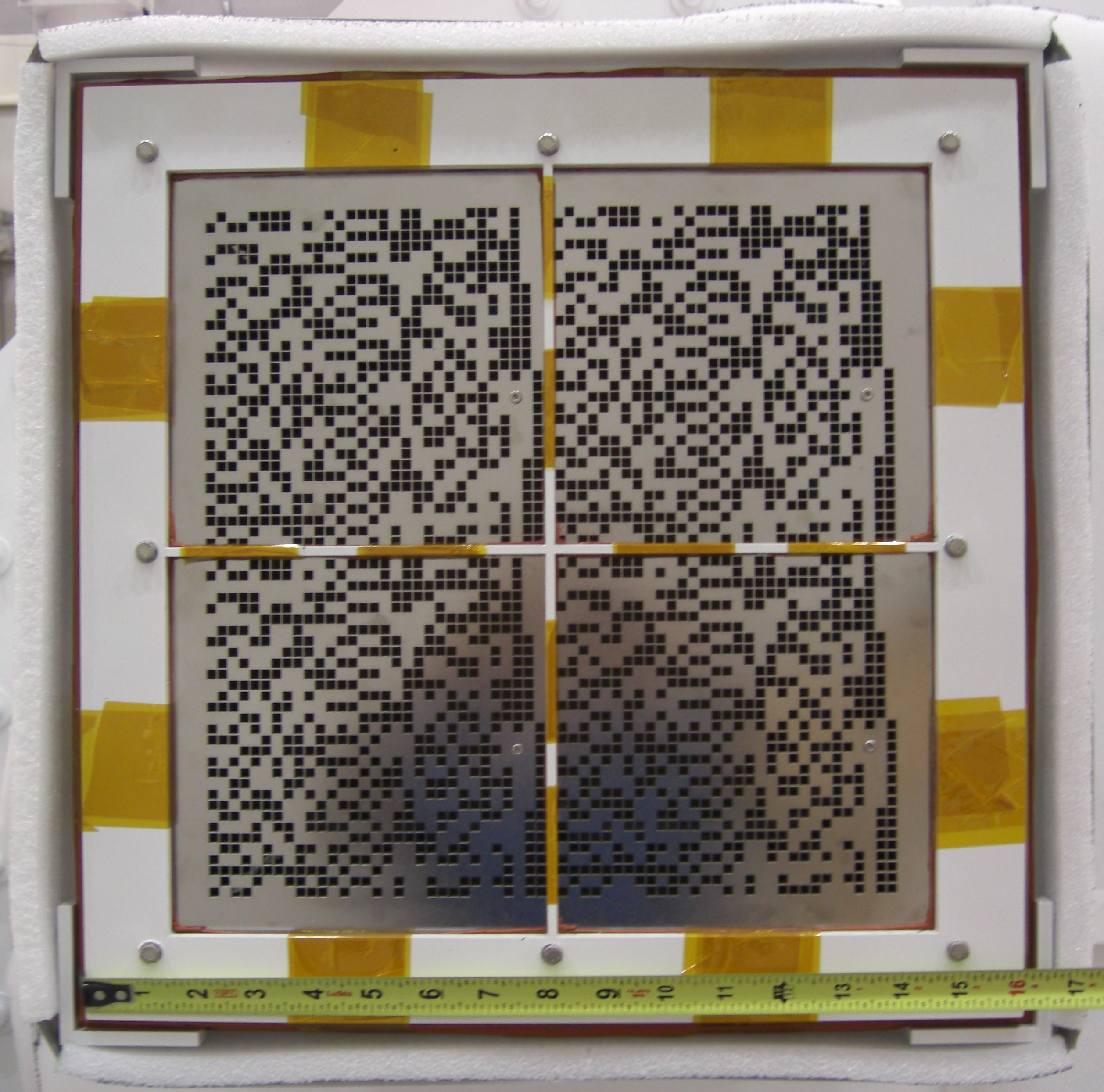}}
	\subfloat[]{\label{sfig:dimg}\includegraphics[width=0.29\textwidth]{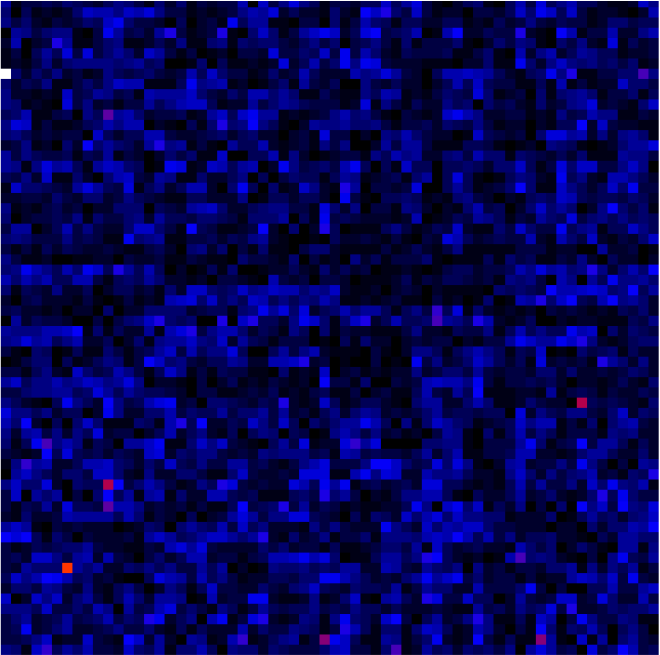}}
	\subfloat[]{\label{sfig:rimg}\includegraphics[width=0.29\textwidth]{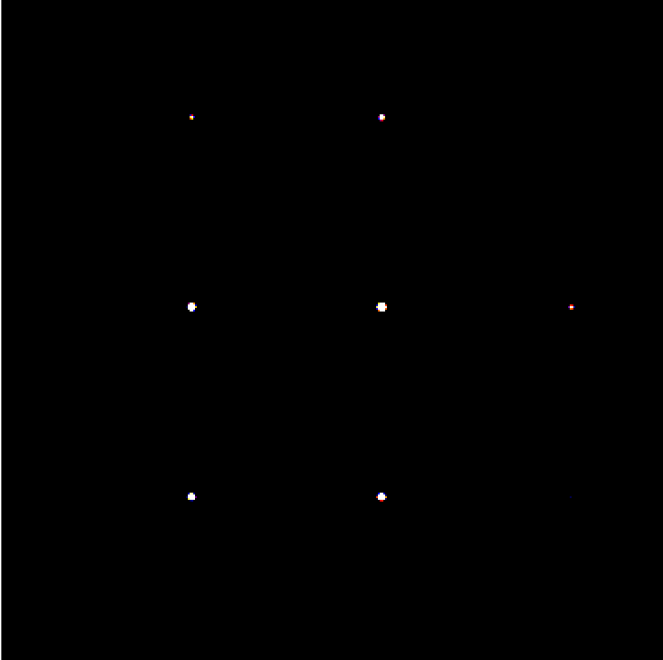}}
	\caption{(a) The {\it ProtoEXIST1} URA mask and support tower mounted to the top 
	surface of the pressure vessel prior to flight.  Each pixel is 4.2 mm square and is 
	separated by a 0.5 mm thick support grid. Prior to flight an imaging test was carried
	out using a 3 mCi $^{57}$Co source placed approximately 3 m from the surface of the 
	mask. (b) The mask pattern can be seen in the detector plane image generated from the
	data collected during the imaging test.  (c) The reconstructed image produced 
	from a FFT cross correlation between the mask pattern and the raw detector image 
	clearly reconstructs the source.  Due to the cyclic nature of the URA 8 "ghost" images
	are also visible.}
	\label{fig:mask}
\end{figure}
The final assembled telescope has a $10^\circ\times10^\circ$ fully coded field of view
and an angular resolution of $21^\prime$.

Prior to the construction of the mask a series of imaging simulations were carried out
for optimization of the mask pattern.  10 separate masks were simulated utilizing 
the current URA pattern with different scale sizes between 4.5 mm and 5.5 mm 
at equal 0.1 mm steps within this range.  In each case the 0.5 mm support grid was retained and 
the mask pixel size was reduced to achieve the change in pixel pitch.  The detector plane 
was modeled to represent the real detector plane as closely as possible with 600 $\mu$m gaps 
between each DCU.  Due to the structure of the detector plane and mask additional induced 
coding noise was predicted to occur in detector images generated with the planned 
5.0 mm mask.  From these simulations it was determined that a mask with a 4.7 mm pitch
minimized this effect and was ultimately adopted for use in the 
{\it ProtoEXIST1} telescope.  An imaging verification test was carried out with the 
fully integrated telescope prior to flight utilizing a single 
3 mCi $^{57}$Co source placed at an approximate distance of 3.0 m from the surface of 
the mask.  The performance of the test image was consistent with that of the simulated 
images after taking the magnification factor due to the placement of the source 
at a finite distance into account (Allen {\it et. al.} in preparation\cite{2010FUTU...00...000A}).


\section{Flight Performance}\label{sec:flt}
\begin{figure}
	\centering
	\subfloat[]{\label{sfig:temp}\includegraphics[width=0.45\textwidth]{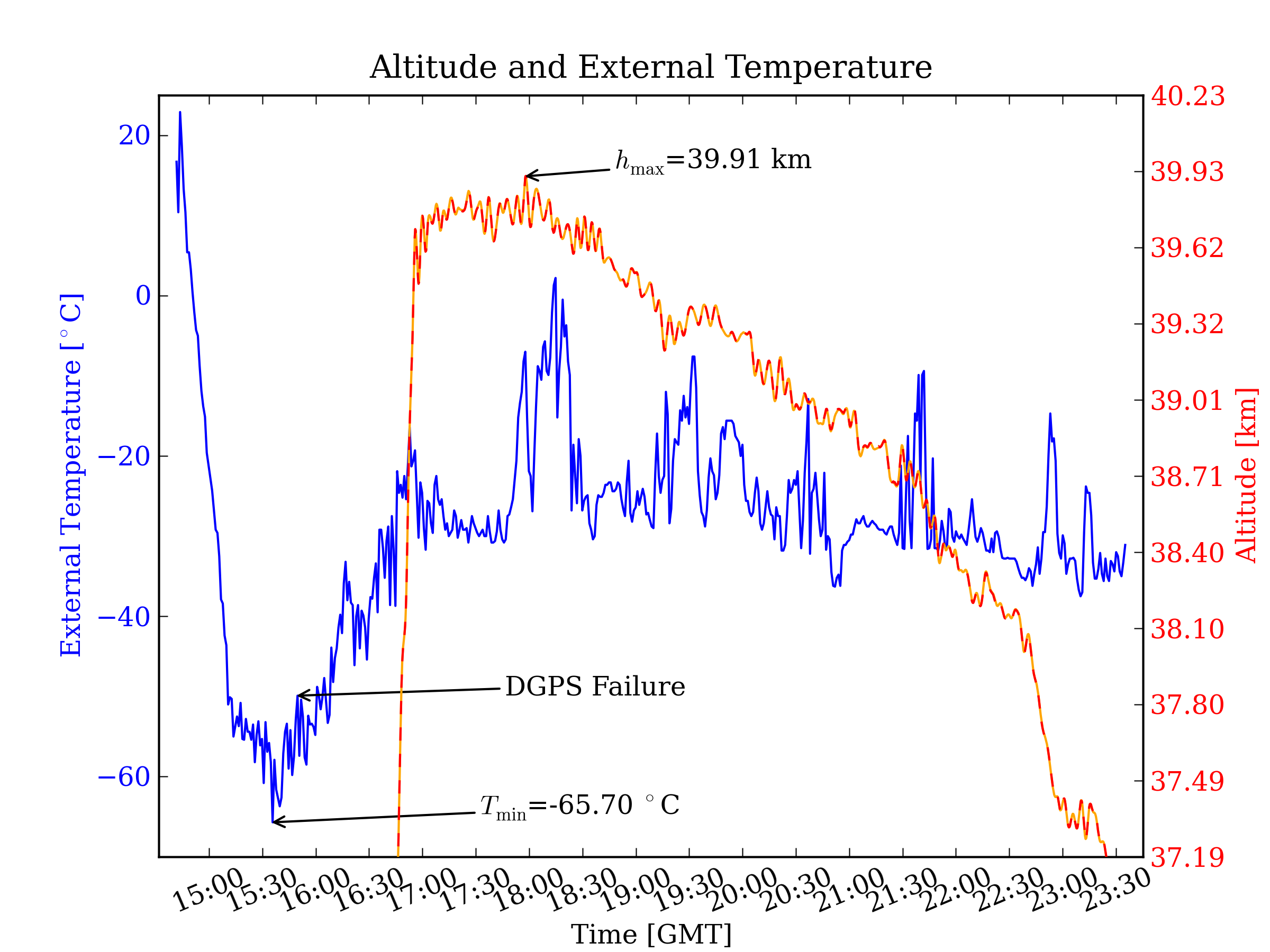}}
	\subfloat[]{\label{sfig:pnt}\includegraphics[width=0.45\textwidth]{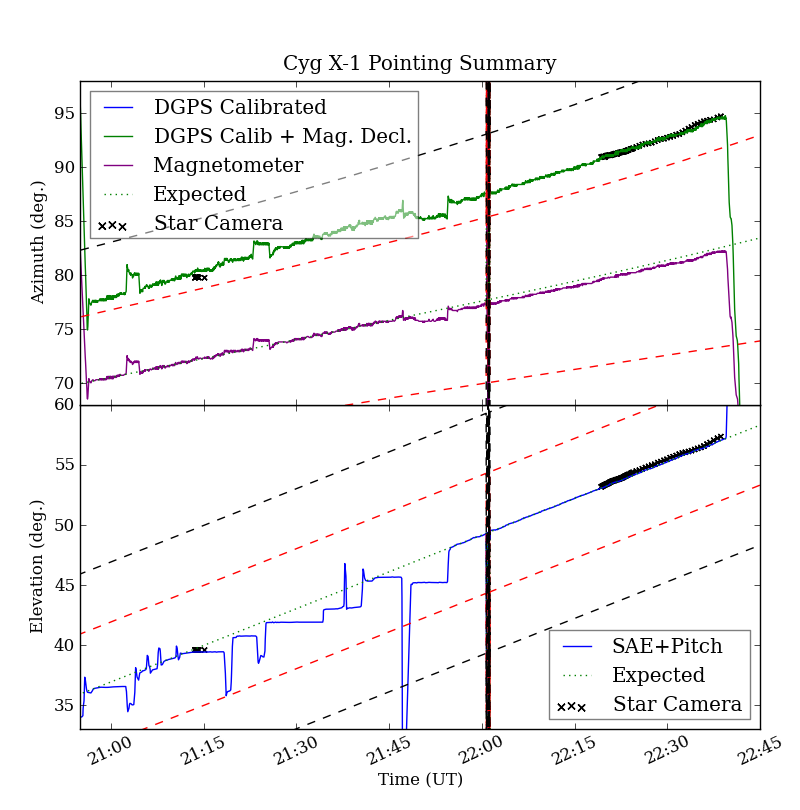}}
	\caption{(a) The temperature and altitude recorded during the flight.  The
	DGPS suffered a catastrophic failure at approximately 15:50 UT (see table \ref{tbl:time}), 
	nearly coincident
	with the observation of the minimum external temperature, forcing the aspect system
	to rely on the backup magnetometer for the determination of the coarse azimuth position.
	(b) The aspect data taken during the Cyg X-1 observation which was carried out using
	only the magnetometer and the shaft angle encoder.  Instability in the elevation axis, and 
	to a limited extent in the azimuth, can clearly be seen prior to 22:50 UT.  Afterward the 
	instability disappeared and stable tracking was maintained until termination of
	observation at approximately 22:40.  The black crosses are the reconstructed aspect solutions
	for each of the available star camera images retrieved after landing which confirm the validity
	of the corrected magnetometer aspect solution.  The red dashed lines and the black dashed lines show 
	the extent of the fully and partially coded field of view respectively. 
	A post-flight analysis of the aspect data revealed a $\sim7^\circ$ offset from the target 
	position, consistent with the reconstructed position of Cyg X-1 and star camera measurements.}
	\label{fig:pnt}
\end{figure}

{\it ProtoEXIST1} began its maiden flight at 14:40 UT on the 9$^\mathrm{th}$ 
of October 2009.  The gondola reached float altitude (nominally 39 km) 
approximately 1.5 hours after liftoff (see figure \ref{sfig:temp} and table \ref{tbl:time}).
The observation plan was initiated with an attempted observation of the quasar 3C 273, 
however during the first half of the flight due to instability in the elevation axis 
a stable pointing count not be achieved, presumably due to a stiff wire on the elevation 
axis.  After the end of the 3C 273 observation run testing of the CsI rear shield thresholds was 
carried out over a period of approximately 30 min.  At 15:00 an observation of 
the black hole binary Cyg X-1 was initiated and carried out over the course of 
one hour.  A short system power control test and detector restart was initiated after
which time the Cyg X-1 observing run was re-initiated until flight termination.
Cyg X-1 was successfully observed and the payload was recovered with minimal damage
to the gondola.  The science package was recovered unharmed.

In spite of the short duration of the flight and the additional difficulties encountered
with the pointing system the primary objectives: the detection of a celestial source (Cyg X-1)
with the {\it ProtoEXIST1} coded aperture telescope, the successful operation of the 
detector plane and measurement of backgrounds in a near space environment, and verification 
that all the payload subsystems functioned as designed in flight were all accomplished.
\begin{table}
	\centering
	\include{timeline}
	\caption{A brief overview of the milestones which occured throughout the flight.}
	\label{tbl:time}
\end{table}

\subsection{Pointing System Performance}\label{ssec:pnt_sys_per}
During the flight 3 major difficulties with the pointing system were encountered which 
complicated flight operations and the analysis of the flight data: the failure of 
the differential global positioning system (DGPS) shortly after liftoff, the failure 
of the daytime star camera to acquire an aspect solution during the flight, and sporadic instability 
in the elevation axis.
\begin{figure}
	\centering
	\includegraphics[width=\textwidth]{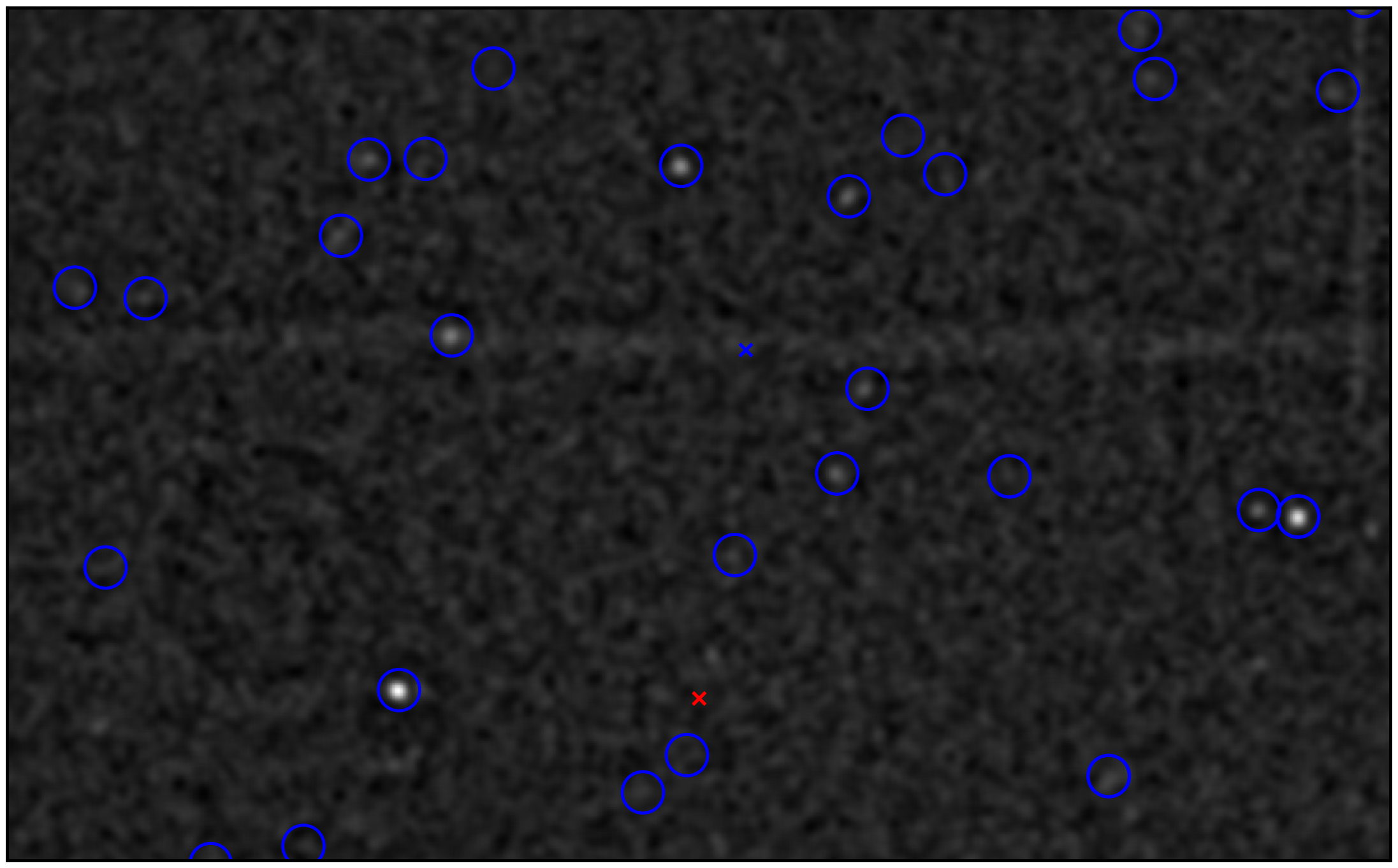}
	\caption{A single star camera image taken at 22:22:19 UT on the 9$^\mathrm{th}$ 
	of October 2009.  In order to extract sources from a single star camera image 
	a large scale background subtraction is applied using a boxcar smoothed image 
	with a $75\times75$ pixel kernel as a rough background estimate.  A 2D gaussian filter 
	with $\sigma=5$ pixels is then applied and a source list extracted with 
	SExtractor\cite{1996A&AS..117..393B}.  The source list is passed to 
	astrometry.net-0.25\cite{2010AJ....139.1782L} in order to determine an aspect 
	solution for the image.  The post processed image is shown above with the detected 
	sources marked with blue circles.  The stars detected in this image have a 
	$V_\mathrm{mag}$ between approximately 9.5 and 9.0.}
	\label{fig:star_camera}
\end{figure}

During the first half of the flight instability of the elevation axis was a major issue
(see figure \ref{sfig:pnt}) and prevented the stable observation of 3C 273.  
Fortunately stability improved as the flight progressed and a long observation 
of Cyg X-1 was possible.  The suspected
cause is a combination of stiff wiring at low temperature in combination with a
relatively weak elevation axis motor.  In anticipation of future flights the replacement
of the elevation axis motor is being considered.

The most serious issue encountered was the loss of the DGPS module during the ascent 
of the payload (see figure \ref{sfig:temp}).  As the DGPS is the primary method by which 
an accurate azimuthal position for the coarse pointing system is determined its loss forced
the aspect system to fall back on the magnetometer.  Unfortunately magnetometers have large
intrinsic systematics associated with their orientation relative to the position of other
structures on the payload as well as their position on the earth;  a post-flight 
analysis of the magnetometer data revealed substantial systematic offsets of up to 
$50^\circ$. On ascent attitude control was offline and the gondola was allowed to 
rotate freely, which enabled the collection of simultaneous measurements 
of the azimuth by both the magnetometer and the DGPS over the entire 360$^\circ$.  These 
measurements were later used to calibrate the magnetometer, which removed the most egregious 
systematic errors in the azimuth measurement.  After the application of an additional
correction taking into account the change of the magnetic declination over the course 
of the flight
utilizing GEOMAG IGRF10\footnote[1]{http://www.ngdc.noaa.gov/IAGA/vmod}, a model of the 
earth's magnetic field, the pointing data was reconciled with aspect solutions determined 
from archived star camera images.  The cause of the DGPS
failure is currently under investigation, however, since the DGPS unit is mounted externally and
the failure was nearly coincident with the observation of the minimum recorded external 
temperature it is highly likely it was the result of insufficient thermal insulation and 
heaters.  To guard against failure in future flights new heating and insulation schemes 
are under consideration.
As an additional safeguard an on-board magnetometer calibration subroutine is also being
contemplated to mitigate the large systematic pointing errors in the event of a DGPS failure
on the next flight.

The next major difficulty was the failure of the daytime star camera to 
acquire a pointing solution during the flight.  This was primarily the result of
bad columns in the star camera CCD, a large offset between the true pointing position
and the measure position due to the loss of the DGPS, and to a lesser extent insufficient 
shielding from the atmospheric background light at certain positions on the sky.  The
offset of the measured azimuthal value using the magnetometer likely prevented the determination 
of an aspect solution during the majority of the flight due to the finite star catalog search 
radius (approximately $4^\circ$) about the coarse aspect system's reported position.  
To correct for these problems the CCD has been returned to the manufacturer for repair, 
an evaluation of the refurbished CCD is planned within the next year.  In addition to this
alternative star field search algorithms are being considered for the star camera in 
order to increase the size of the search radius.  In spite of these difficulties 
usable star camera images archived during the flight proved useful in reconstruction of 
the telescope aspect information.  Using cleaned archived star camera
images the extraction of star positions in the image using SExtractor\cite{1996A&AS..117..393B} 
then an aspect solution using astrometry.net-0.25\cite{2010AJ....139.1782L} was possible.
Images were cleaned by removal of bad columns from the
image and replacement of the bad column pixel values with the average taken from 
their 6 nearest neighbors, a rough flat fielding and application of a 2D Gaussian 
filter with $\sigma=5$ pixels was applied to the entire image 
(see figure \ref{fig:star_camera}).  After the aspect information for all star camera images
was determined an aspect solution was generated then compared to the corrected magnetometer /
elevation aspect data.  Using this it was possible to independently confirm the validity 
of the corrected magnetometer aspect solution (see figure \ref{fig:pnt}).

\subsection{Telescope System Performance}\label{ssec:tel_sys_per}
The telescope and the associated subsystems performed without incident during the
flight.  The science payload heating system and control mechanism maintained a 
steady internal temperature of $20\pm3^\circ$C for the entire duration of the flight.
The flight computer and associated command and control system did not experience a single
failure enabling the receipt of data from liftoff through flight termination.
The internal power control and monitoring mechanism proved adequate; toward the end of the 
flight a system test was conduced to confirm the stable operation of the independent 
CIP interfaced power control mechanism as well as to test the complete remote 
reinitialization of the detector plane and other subsystems.  The restart procedure, save a 
minor glitch in the data readouts, was a complete success.  Within 15 min. after the restoration
of power the telescope and all subsystems were running within nominal parameters and celestial 
observations were resumed.

\begin{figure}
	\centering
	\subfloat[]{\label{sfig:edge_bkg}
			\includegraphics[width=0.49\textwidth]{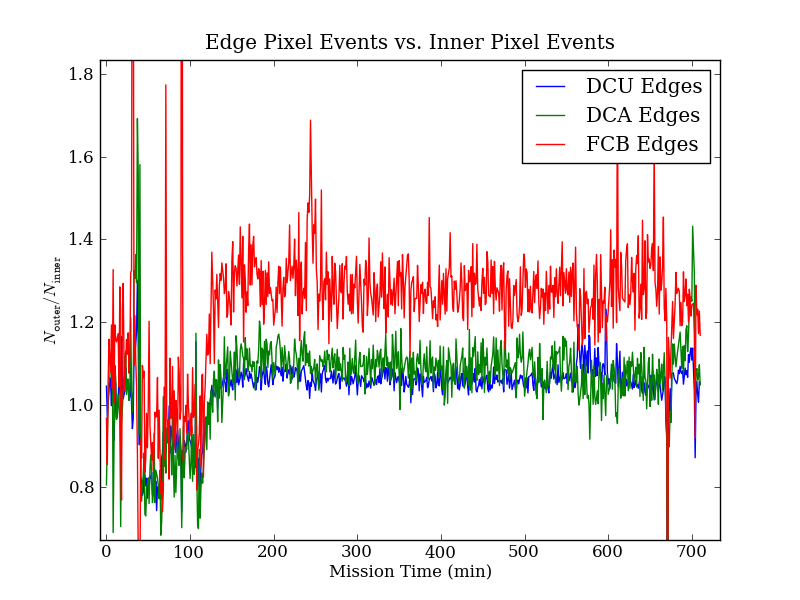}}
	\subfloat[]{\label{sfig:bkg_rate}\includegraphics[width=0.49\textwidth]{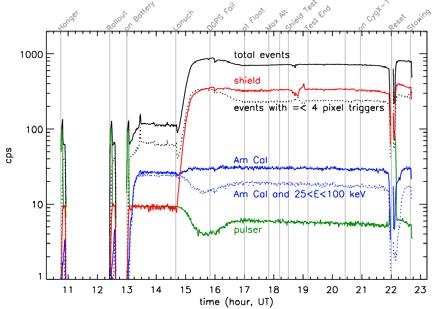}}
	\caption{(a) The ratio of the total background rate for pixels located on the 
	edge of the FCB (red), on the edges between the individual DCAs (green), 
	and on the edges between the individual DCUs (blue), to total the background rate 
	of pixels located within the DCU's.  Pixels located on the periphery of the 
	detector plane experienced a 30\% higher background rate than the pixels located 
	in the middle of the DCUs, while those located on the edges adjacent to neighboring 
	detectors only experienced a 5\% increase. (b) The detector rates over the entire flight
	remained remarkably stable once the payload reached float altitude.  The only changes
	in rate occurred in the shield during the test adjustment of the threshold values
	(around 18:30 UT) and during reinitialization of the detector system after the 
	power test (22:00 UT) and the required change in threshold values on restart.}
	\label{fig:bkg}
\end{figure}

In order to verify the operation of the coded-aperture telescope an observation of Cyg X-1 and
3C 273 was carried out.  As outlined in \S\ref{ssec:pnt_sys_per} the observation of 3C 273 proved
untenable due to issues encountered with the pointing system, however sufficient pointing stability
was achieved during the observation of Cyg X-1.  After application of corrected aspect 
data Cyg X-1 was detected at 7.2$\sigma$ with $\sim7^\circ$ offset after 10 minutes of 
stable observation at the end of the flight (c.f. Hong {\it et. al.} in preparation
\cite{2010FUTU...00...000H}).
%
\begin{figure}
	\centering
	\subfloat{\label{sfig:gspek}\includegraphics[width=0.49\textwidth]{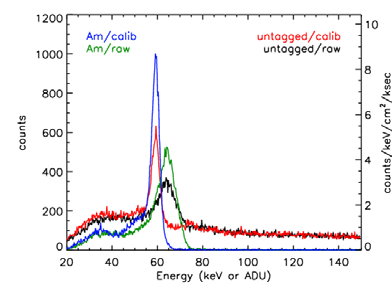}}
	\subfloat{\label{sfig:fspek}\includegraphics[width=0.49\textwidth]{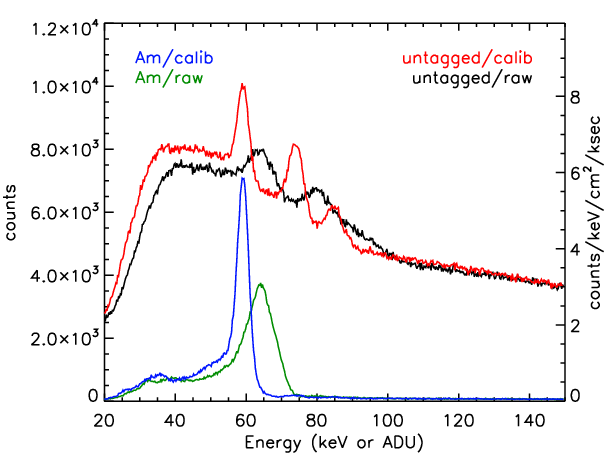}}
	\caption{(a) The preflight calibrated and non-calibrated spectra over the entire detector 
	plane for the $^{241}$Am calibration source and the background spectrum measured in the 
	hanger at Ft. Sumner, NM.  The measured energy resolution after calibration is 3.5 keV FWHM at
	59.6 keV.  (b) Calibrated and uncalibrated, background and calibration spectrum observed in
	flight.  The two additional lines visible above the 59.6 keV $^{241}$Am line are the 
	K$\alpha$ and K$\beta$ fluorescence lines emitted from the segment of unshielded Pb on the
	lower side shield section (see \S\ref{ssec:shld}).  The energy resolution measured in flight
	is approximately 4 keV FWHM at 59.6 keV.}
	\label{fig:spectrum}
\end{figure}
\begin{figure}
	\centering
	\includegraphics[width=0.75\textwidth]{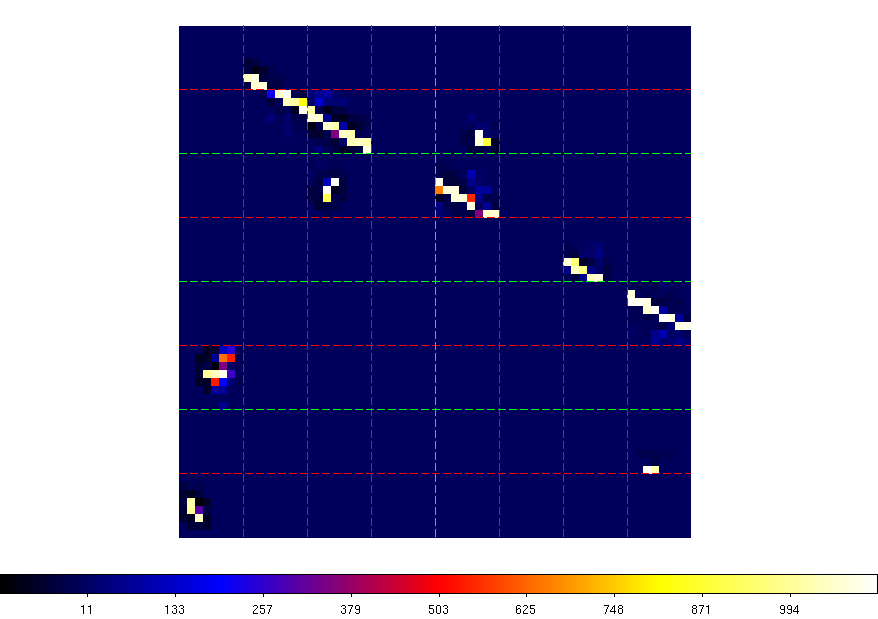}
	\caption{A detector plane image of the track and charge deposited by a single cosmic ray skimming 
	the top of the detector plane.  The breaks in the track are a consequence of the 
	independence of the 32 individual DCU pairs and the $\sim30$ $\mu$s deadtime.  This image
	was generated taking a coincidence window of 200 $\mu$s between events recorded
	on individual DCU pairs.  The color scale is the uncalibrated energy recorded in each
	detector pixel.  The dashed green lines represent the DCA borders, and the dashed red
	lines represent the borders between DCUs.}
	\label{fig:CR}
\end{figure}
During the flight the background spectra and detector behavior were also monitored
for any deviations in behavior due to charge saturation in the high radiation environment.
In spite of the large increase in the background event rate the detector plane
operated normally without any degradation in performance.  The spectral response of the
detector was monitored in flight using the tagged $^{241}$Am source and,
as was observed on the ground the and during the flight.  A slight increase in the 
average energy resolution was observed, 4.0keV in flight vs. 3.5 keV in the hanger (FWHM) 
at 59.6 keV averaged over all active pixels for the duration 
of the flight.  The detector gain remained constant over the duration of the flight as did  
the background at $4.0\times10^{-3}$ cm$^{-2}$ s$^{-1}$ keV$^{-1}$ 
above 120 keV; in line with expectations.  Additionally one of the key advantages
of the tiled detector configuration proposed for {\it EXIST}, namely the reduction
of background rates through shielding of the individual detector edges with a 
neighboring detector, was clearly demonstrated as well.  For pixels on the edge of the 
detector plane, 252 in all, the measured backgrounds were 30\% higher than observed in 
pixels located in the central (i.e. non-edge) pixels of the DCUs. For edges shielded by 
an adjacent detector, 1540 pixels total, a $\sim5\%$ higher background rate was observed.  
Extrapolating from these rates had the {\it ProtoEXIST1} detector not been closely tiled 
fully exposing all DCU edges a 10\% increase in the background rate would be expected 
(see figure \ref{fig:bkg}).  

\section{Future Telescope Development}\label{sec:future}
Development for {\it EXIST} is currently continuing with {\it ProtoEXIST2}.
{\it ProtoEXIST2} will utilize the DB-ASIC for the readout of 
19.5 mm $\times$ 19.5 mm $\times$ 5.0 mm CZT detectors with a 
$32\times32$ array of anode pixels with a 600 $\mu$m pitch.  Using 
a slightly modified version of the {\it ProtoEXIST1} modularization 
scheme (see \S\ref{sec:det}); a $4 \times 4$ sub-array of these detectors 
(1/4 of a detector plane) will be fabricated and demonstrated. It is 
anticipated that testing will begin on the first detector modules 
for {\it ProtoEXIST2} within the next few months 
with the completion of the first sub-array occurring before April of 2011.
In {\it ProtoEXIST3} the remaining 3/4 of the detector plane will be 
built using a lower power version derived from the DB-ASIC which 
consumes about a factor of four less power, the {\it EXIST Final} or EX-ASIC.
With the completion of the detector plane a test flight with both the original
{\it ProtoEXIST1} and new {\it ProtoEXIST2/3} detector planes will be
be carried out in the fall of 2012.  The {\it ProtoEXIST2/3} telescope 
will feature an extended 2 m focal length, be capable of imaging sources 
with an angular resolution of $2.4^\prime$ and have a fully coded 
$2.3^\circ\times2.3^\circ$ field of view.  Parallel to this effort a 
further modification of the EX-ASIC design for the replacement of
wirebond pads, which prevent close tiling over the entire detector plane, 
with microvias to form the final {\it EXIST} ASIC (EXF-ASIC).

\section{Conclusions and Future Plans}\label{sec:conclusion}
The successful integration of the {\it ProtoEXIST1} detector plane and its operation
at high altitude in a near space environment have demonstrated the feasibility 
of operating a large area tiled detector plane similar to the {\it EXIST-HET} 
utilizing individual pixiliated CZT detectors (DCUs) as the core building block.
Measured spectra revealed that the high background rates and charge saturation due
to incident cosmic rays had little impact ($\sim10\%$) on the detector energy resolution
and no discernible impact on the detector gains.  The elevated event rates 
also did not pose any difficulties for the detector plane electronics or the 
data acquisition system. It has also been demonstrated that the interference 
between the individual DCUs on the ground and in flight does not pose a 
significant challenge to the construction of large tiled arrays. 
In short the operation of the telescope and science package was a complete success
on the ground and in flight. Additionally a key advantage of the tiled array, 
the mutual shielding of detector edges by neighboring DCUs, was also demonstrated
in flight showing that a reduction in backgrounds of at least 10\% is possible 
over a non-tiled detector plane configuration.

In the near future repair of the landing damage on the gondola will be carried out
as will improvements to the pointing system to improve reliability.  Once the detector
integration for {\it ProtoEXIST2/3} is completed a flight together with the 
original {\it ProtoEXIST1} detector plane is planned for characterization and comparison 
of both systems in flight.

\bibliography{ref,future}
\bibliographystyle{spiebib}   
\end{document}